\documentclass[10pt,a4paper,twoside]{article}

\usepackage{indentfirst}            
\usepackage{graphicx}               
\usepackage{amstext,amsgen,amsfonts,amssymb,amsbsy} 
\usepackage{xspace}

\newenvironment{resum}{\begin{quote}\small}{\end{quote}}

\newcommand{\bfsl}[1]{\textsl{\textbf{#1}\/}}

\newcommand{\bma}[1]{\mbox{\boldmath${#1}\/$}}

\newcommand{\pten}[1]{10\mbox{$^{#1}$}}
\newcommand{\fparen}[2]{\left(\frac{#1}{#2}\right)}
\newcommand{\eq}[1]{(\ref{#1})}

\newcommand{\x}{\text{\kern.1em$\times$\kern.1em}}
\newcommand{\bOx}{\raisebox{-0.6 pt}{\large $\Box$}}

\newcommand{\esa}{\textsl{ESA}\xspace}
\newcommand{\lisa}{\textsl{LISA}\xspace}
\newcommand{\lpf}{\textsl{LPF}\xspace}
\newcommand{\ltp}{\textsl{LTP}\xspace}
\newcommand{\nasa}{\textsl{NASA}\xspace}

\begin{document}

\thispagestyle{plain}       

\begin{center}


{\LARGE\bfsl{LISA}}

\bigskip


\textbf{J.\ Alberto Lobo}


\emph{Universitat de Barcelona \& Institut d'Estudis Espacials de Catalunya}

\end{center}

\medskip


\begin{resum}
The extreme weakness of the gravitational interaction has as one of
its consequences that appreciable intensities of gravitational waves
(GW) can only be generated in large size astrophysical and cosmological
sources. Earth based detectors face unsurmountable problems to be sensitive
to signals at frequencies below 10 Hz due to seismic vibrations. In order
to see lower frequency signals, a space based detector is the natural
solution. \lisa (Laser Interferometer Space Antenna) is a joint \esa-\nasa
project aimed at detecting GWs in a range of frequencies between
\pten{-4} Hz and \pten{-1} Hz, and consists in a constellation of three
spacecraft in heliocentric orbit, whose GW-induced armlength variations
are monitored by high precision interferometry. This article reviews
the main features and scientific goals of the \lisa mission, as well
as a shorter description of its precursor tecnological mission \lpf
(\lisa Pathfinder).
\end{resum}

\bigskip


\section{Introduction}

The detection of gravitational waves (GW) is one of the most challenging
scientific problems ever in the history of science. Ultimately, this is
due to the extreme weakness of the gravitational interaction ---40~orders
of magnitude weaker than electromagnetism.

The nature and properties of gravitational radiation were first described
by Einstein shortly after he created the General Theory of Relativity
\cite{lobo-ein18}. However, beyond the specific predictions of Einstein's
theory, the phenomenon of GWs appears to be an unavoidable consequence
of the basic fact that no interaction can travel instantaneously from
source to observer. Recall e.g.\ that Poisson's equation for the
gravitational potential of Newtonian gravity,
\begin{equation}
 \nabla^2\phi({\bf x},t) = -4\pi G\,\varrho({\bf x},t)\;,
 \label{lobo-eq1.1}
\end{equation}
where $\varrho({\bf x},t)$ is the density of matter creating the
gravitational field, strongly violates such basic causality principle,
and therefore needs a major conceptual reassessment. Whether this is
provided by Einstein's theory or any of its competitors ---for
example Brans--Dicke theory~\cite{lobo-bd61}, see also~\cite{lobo-will}---
GWs should be there, anyway. Actually, a GW detection experiment
is a potential tool to decide between candidate theories.

Gravitational waves are characterised mathematically by dimensionless
amplitudes $h_{\mu\nu}(\bma{x},t)$, measuring deviations from a flat
space-time geometry caused by the radiation field. Far from their
sources, these amplitudes are usually very small, so the metric
tensor $g_{\mu\nu}(\bma{x},t)$ splits up as
\begin{equation}
 g_{\mu\nu}(\bma{x},t) = \eta_{\mu\nu} + h_{\mu\nu}(\bma{x},t)
 \label{lobo-eq1.2}
\end{equation}
in a quasi-Lorentzian coordinate frame, i.e.,
\begin{equation}
 \eta_{\mu\nu} = {\rm diag}\;(-1,1,1,1)\ ,\qquad
 \left|h_{\mu\nu}(\bma{x},t)\right|\ll 1\;.
 \label{lobo-eq1.3}
\end{equation}


Not all ten of the quantities $h_{\mu\nu}(\bma{x},t)$ contain
independent information on the physics of GWs; these, as is well
known from textbooks on General Relativity, have rather only two
canonical degrees of freedom, usually noted $h_+$ and $h_\times$,
after a universally accepted terminology, proposed by Misner,
Thorne and Wheeler~\cite{lobo-mtw}. We recall that, for a GW travelling
down the $z\/$-axis, the $h_{\mu\nu}$'s actually reduce to
\begin{equation}
 h_{\mu\nu}(\bma{x},t) = \left(
 \begin{array}{cccc}
  0 & 0 & 0 & 0 \\
  0 &   h_+(ct-z)        &  h_\times(ct-z) & 0 \\
  0 & \ h_\times(ct-z)\  & \ -h_+(ct-z)\   & 0 \\
  0 & 0 & 0 & 0
 \end{array}\right)
 \label{lobo-eq1.35}
\end{equation}

Using the quadrupole approximation of General Relativity~\cite{lobo-wein},
one can find a crude order of magnitude estimate of the intensity of
GWs far from their sources; it is given by the formula~\cite{lobo-sussp}
\begin{equation}
  h \sim \frac{R}{r}\,\frac{v^2}{c^2}
  \label{lobo-1.4}
\end{equation}
where $R\/$\,=\,$2GM/c^2$ is the Schwarrzschild radius of the source,
$v\/$ is a typical velocity of the matter inside it, and $r\/$ is the
distance to the observer, assumed orders of magnitude larger than $R$.
The characteristic frequency of emission is on the other hand of order
$v/a$, where $a\/$ is roughly the linear size of the source.

\begin{figure}[t]
\centering
\includegraphics[width=11.5cm]{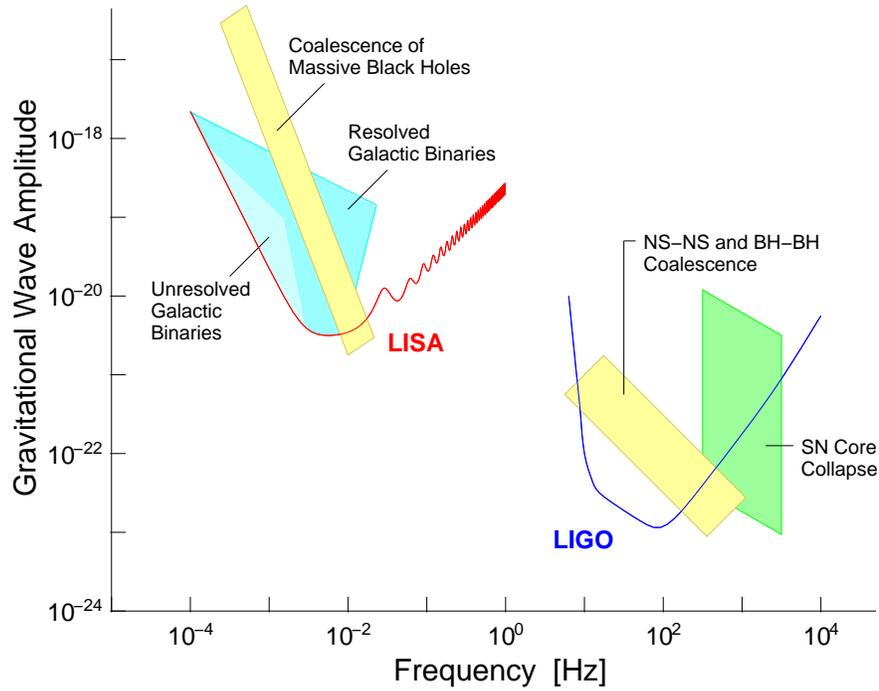}
\caption{Spectral densities of noise of two very different GW detectors,
{\sl LIGO\/} and \lisa, along with some of the signals they can sense.
\label{lobo-fig2}}
\end{figure}

The range of frequencies in which GWs happen can widely vary from
\pten{-18}~Hz, corresponding to primeval GWs, up to \pten{4}~Hz or
higher, corresponding to small mass BH oscillation modes.
Amplitudes are however less variable, and they are very small when
received on Earth, even in the best of cases. Taking into consideration
the large uncertainties in their estimation, current common lore sets
the amplitude range of GWs reaching the solar system somewhere in the
interval
\begin{equation}
 \pten{-26}\lesssim h \lesssim\pten{-16}
  \label{lobo-1.5}
\end{equation}

A convenient characterisation of GW signals is in terms of their
\emph{frequency spectrum}. The relevance of this comes from the fact
that any given detector has its own \emph{frequency window}, which
determines in turn which sources it will be able to sight.
Figure~\ref{lobo-fig2} shows a very sketchy summary of source spectra,
together with \emph{noise spectral densities} of two interferometric
detectors which will eventually see the corresponding signals. These
detectors are {\sl LIGO\/}, a ground based device, and \lisa, a space
borne GW telescope. A more complete description of sources is given
in~\cite{lobo-bfs99}. The most salient feature of the graph in figure~\ref{lobo-fig2}
is the \emph{position in frequency} of the sensitivity curves of both
detectors, separated by 4--5 orders of magnitude.

As already stressed, detailed estimation of source amplitudes is
plagued with uncertainties, due to the complicated dynamics of
astrophysical systems. The \emph{cleanest} systems are black hole
binaries, which makes of them excellent candidates for observation.
But uncertainties here are also significant, as we only have a few
hints on their abundances. On general grounds, the following argument
should work: since interesting GWs require large mass motions to be
generated, the most intense sources will likely have low frequencies,
as characteristic lengths in large mass distributions should be large,
and typical time scales should therefore also be large, hence frequencies
low.

To have a more quantitative idea of the orders of magnitude in this
somewhat vague line of argument, let us consider the GWs emitted by the
Hulse--Taylor binary pulsar PSR\,1913+16~\cite{lobo-wt89}: its orbital
period is nearly 8 hours, which means it radiates GWs at a frequency of
7$\x$\pten{-5}~Hz. We shall agree that this a is rather low frequency.
On the other hand, we consider that 1 kHz is a high frequency, and
signals around this value are accessible with current ultracryogenic
bars~\cite{lobo-naut,lobo-auriga} and upcoming ground based
interferometers~\cite{lobo-ligo,lobo-virgo}.

The above figures are useful to make sense of what \lisa is intended for:
the search for low frequency GW signals in the range of \pten{-4}~Hz to
\pten{-1}~Hz. This article will describe the reasons for this choice
and the details of how \lisa will be implemented. We shall first review
a few basic facts about GW detection theory and then show how they apply
to \lisa; its sensitivity and detectable sources will also be mentioned.
The technological difficulties to build a detector with the requirements
of \lisa have motivated the launch of a precursor mission, \lpf (\lisa
{\sl Pathfinder\/}), and we shall also explain its objectives and how they
can be accomplished. Spain has recently acquired an active role in the space
borne GW detector project, whose status will be briefly summarised in the
last section.

\section{GW detection concepts}

The weakness of the gravitational interaction shows up in a particularly
dramatic form when it comes to evaluating e.g.\ the \emph{cross section}
for the absorption of GW energy by matter. If one takes for instance a
solid sphere of aluminum, it can be shown~\cite{lobo-clo} that the optimum
cross section is
\begin{equation}
 \sigma\simeq\pten{-24}\,S
 \label{lobo-2.1}
\end{equation}
i.e., 24 orders of magnitude smaller than the transversal area of the
target body\ldots\ This is not worse than the intensity of the interaction
with test electromagnetic fields, see~\cite{lobo-cqg}.

While this is at the root of the difficulties to build a working GW
detector, it also has a positive counterpart: GWs are so weakly absorbed
that they carry undistorted information on the \emph{interiors} of the
sources producing them, therefore making of GWs a unique tool for the
study of such sources~\cite{lobo-bfs99}.

\begin{figure}[t!]
\centering
\includegraphics[width=7cm]{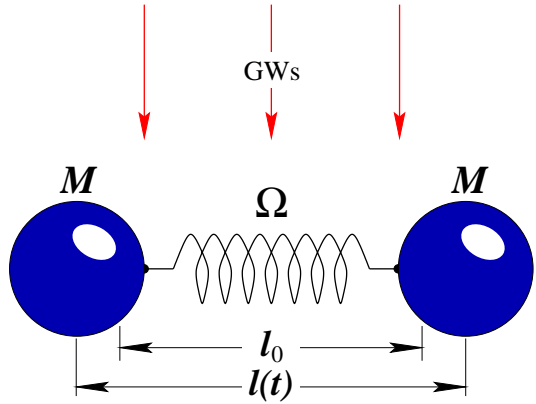}
\caption{The acoustic detector concept: an incoming GW drives the
masses at the spring ends to oscillate, thereby resonantly amplified
motion of the masses is obtained at the spring's characteristic frequency,
$\Omega$.
\label{lobo-fig3}}
\end{figure}

There are at the moment two major GW detector concepts: acoustic and
interferometric detectors. The first is based on the idea of resonant
amplification of the signal, see figure~\ref{lobo-fig3}: two test masses $M\/$
are linked together by a spring of relaxed length $\ell_0$, so that GW
\emph{tides} drive their oscillations around the equilibrium position,
with significant mechanical amplification at the spring's characteristic
frequency $\Omega$. The spring deformation
\begin{equation}
 q(t)\equiv\ell(t) - \ell_0
 \label{lobo-eq2.2}
\end{equation}
thus obeys the following equation of motion\footnote{
Dissipative terms are omitted at this stage, as they do not influence
the key points of our present discussion.}:
\begin{equation}
 \ddot q(t) + \Omega^2 q(t) = \frac{1}{2}\,\ell_0\ddot h(t)\;,
 \label{lobo-eq2.3}
\end{equation}
where
\begin{equation}
 h(t) = \left[
 h_+(t)\,\cos 2\varphi + h_\times(t)\,\sin 2\varphi\right]\,\sin^2\theta\;,
 \label{lobo-eq2.4}
\end{equation}
and $(\theta,\varphi)$ are the angles which define the incidence direction
of the GW in the laboratory frame ---see a complete discussion in
reference~\cite{lobo-pizz}.

This is the main idea, but in practice acoustic GW detectors are
\emph{elastic solids} rather than a single spring, i.e., they do
not have a single characteristic frequency but a whole spectrum.
More complicated analysis tools are needed in this case to find
the detector's response ---the theory is presented in full length
in references~\cite{lobo-lobo,lobo-mnras}. Currently operative acoustic
GW antennas are cylindrical, but spheres are also under consideration
and test~\cite{lobo-grail,lobo-dual}.

\begin{figure}[t!]
\centering
\includegraphics[width=11cm]{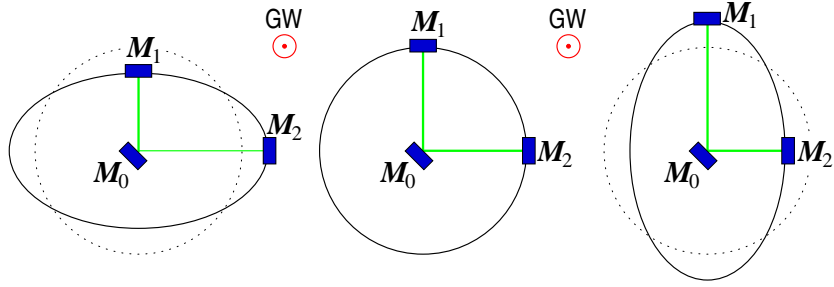}
\caption{The ``na\"\i ve'' interferometric detector concept: a GW coming
perpendicular to the sheet's plane, ``$+$'' polarised relative to the
$x\/$ and $y\/$ axes, causes the end masses $M_1$ and $M_2$ to oscillate
in phase opposition relative to the central mass $M_0$. Light is shone
into the system, and suitable beam splitters and mirrors are attached
to the masses; length changes are then measured interferometrically,
which directly lead to determine the GW amplitude.
\label{lobo-fig4}}
\end{figure}

The second concept for GW detection is interferometry. The idea is
summarised in figure~\ref{lobo-fig4}. In order to honour theoretical rigor,
a warning must be issued regarding this figure: what the interferometer
really measures is a \emph{phase shift} in the light travelling along
two different directions, and this can be determined by solving
Maxwell's equations for a test electromagnetic travelling wave in
the background geometry of a GW, i.e.,
\begin{equation}
\bOx A_\mu\simeq\eta^{\rho\sigma}\,(\partial_\rho\partial_\sigma A_\mu
 - 2\Gamma_{\mu\sigma}^\nu\,\partial_\rho A_\nu
 - \Gamma_{\rho\sigma}^\nu\,\partial_\nu A_\mu
 - A_\nu\,\partial_\rho\Gamma_{\mu\sigma}^\nu) = 0\;,
 \label{lobo-eq2.5}
\end{equation}
where $\bOx\equiv g^{\rho\sigma}\,\nabla_\rho\nabla_\sigma$, and $\nabla$
is the covariant derivative in a geometry given by~(\ref{lobo-eq1.2}).

The phase shift induced on a light beam travelling back and forth in a
Michelson type interferometer of armlength $L\/$ can be found solving
equations~(\ref{lobo-eq2.5}), and the result is~\cite{lobo-cqg}
\begin{equation}
 \delta\phi = \frac{2\omega}{\Omega}\,h_{0}\sin\frac{\Omega\tau}{2}\;,
 \label{lobo-eq2.6}
\end{equation}
assuming a monochromatic, ``$+$'' polarised GW of amplitude
$h(t)=h_{0}\,\sin\Omega t$, which comes in perpendicular to the
inteferometer plane\footnote{Take
$\theta$\,=\,$\pi/2$ and $\varphi$\,=\,0 in equation
(\protect\ref{lobo-eq2.4})}.
Here, $\omega$ is the frequency of the light, and $\tau$ the trip
time of the light in one arm, $\tau=L/c$. Formula~(\ref{lobo-eq2.6})
can be generalised to include arbitrary incidence directions and GW
polarisations~\cite{lobo-cqg,lobo-fara}, but the essential physics of the
interferometer detector concept are already visible in this equation.

More specifically, note that, for a given GW wavelength
$\lambda_{\rm GW}$\,=\,$c/(2\pi\Omega)$, the phase shift is a periodic
function of the the round time $\tau$. The interferometer response has
thus a maximum when its armlength is $L\/$\,=\,$\lambda_{\rm GW}/2$.

A GW of 1~kHz requires an interferometer 150~km long for correct tuning,
and this is the aim of earth based detectors such as {\sl VIRGO\/} or
{\sl LIGO}. These detectors are not 150~km long, but have implemented
an equivalent length by a multiple reflection procedure in the mirrors
whereby light beams actually interfere after they have travelled the
arms back and forth a suitable number of times. Their actual armlength
is only 3~or~4~km.

\subsection{GW detector sensitivities}
\label{lobo-sec2.1}

The above considerations refer only to fundamental properties of GWs
and their interactions with light beams and matter. But a real detector
behaves according to no less fundamental laws regarding the structure
of matter and light. These include the presence of \emph{noise} sources
which unavoidably degrade their sensitivity, and set limits on the
intensity of the GWs they can possibly see.

\begin{figure}[t!]
\centering
\includegraphics[width=6.2cm]{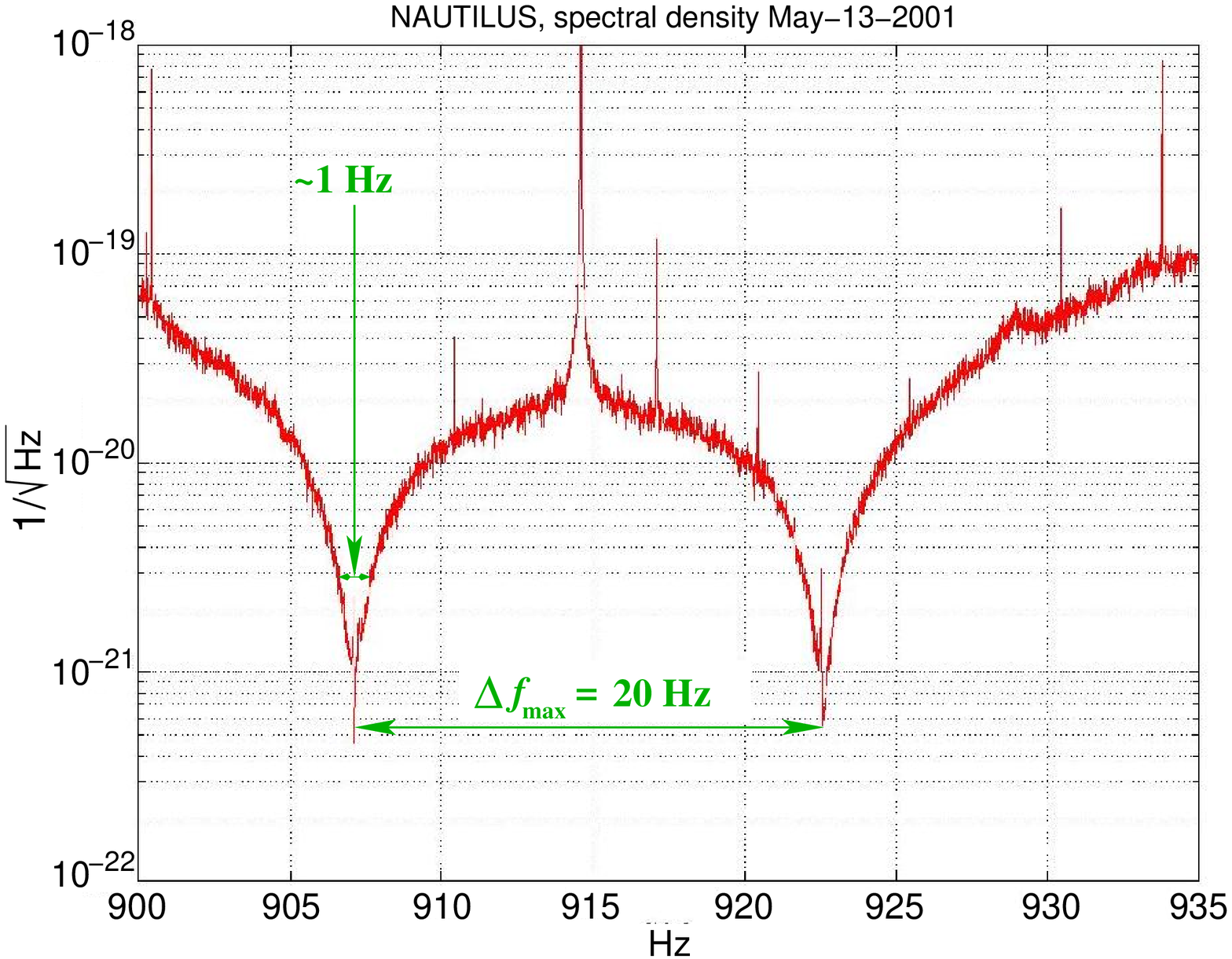}\hspace{1 em}
\includegraphics[width=6.2cm]{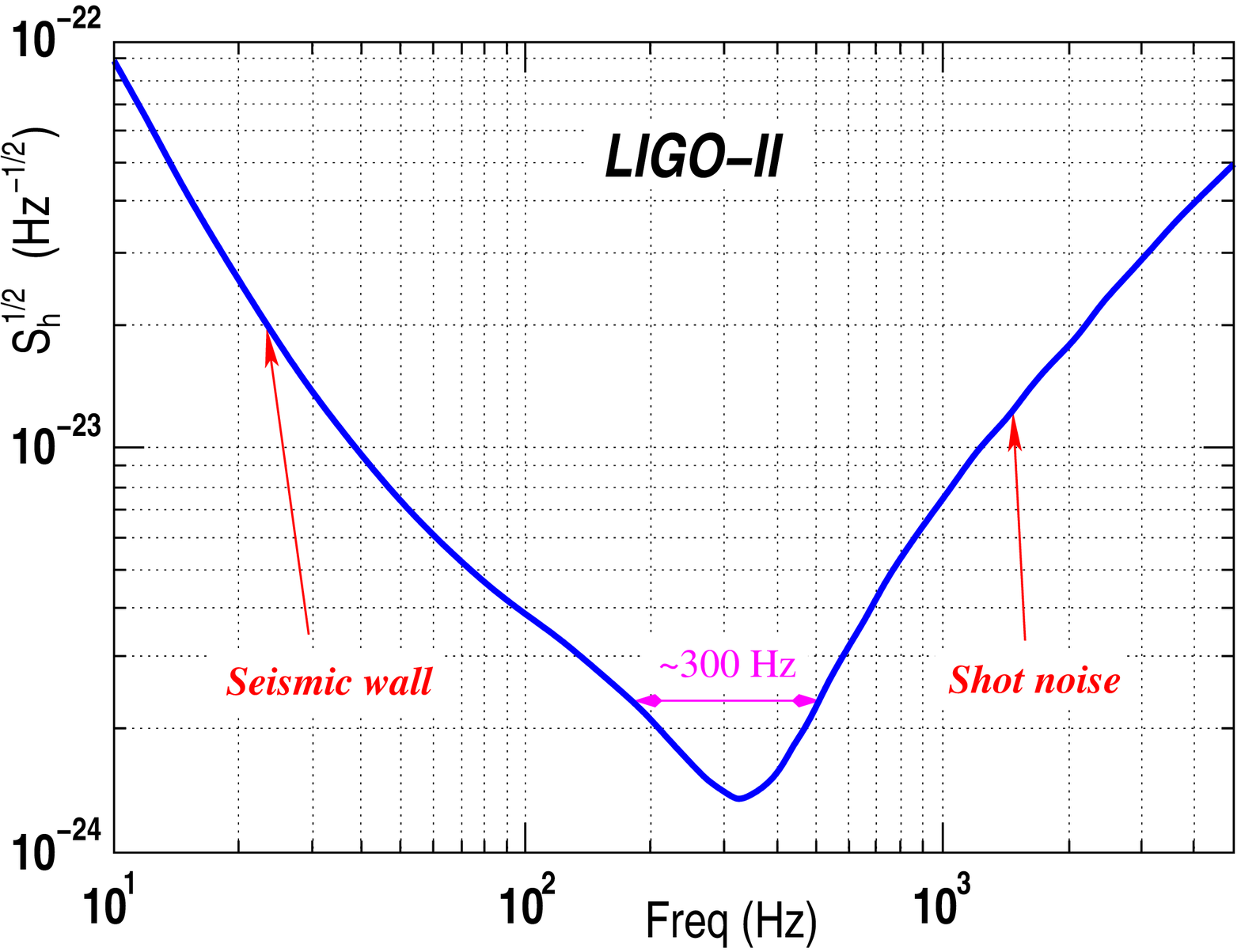}
\caption{Spectral density of noise for two GW detectors: in the left
the narrowband device {\sl NAUTILUS}, and the right the broader band
system {\sl LIGO\/} in its advanced design previsions.
\label{lobo-fig5}}
\end{figure}

By way of example, the plots in figure \ref{lobo-fig5} show the sensitivities,
expressed in spectral density of noise, of two very different GW
detectors: one is the ultracryogenic bar {\sl NAUTILUS}, sitting in
{\sl INFN\/} Laboratories in Frascati, near Rome, and the other is the
long baseline (4~km) {\sl LIGO}. The first is a narrowband device, as
can be seen in the plot, which has two resonant frequencies\footnote{
This is due to the readout system, based on a \emph{resonant transducer},
which causes a split up of the fundamental mode of the cylinder, otherwise
in the middle of the two lines seen.}. Due to fundamental reasons
\cite{lobo-paik}, The best bandwidth which can be possibly attained with
this device is some 20~Hz, the distance between the two resonant peaks,
as also shown in the graph. On the other hand, see right panel in
figure~\ref{lobo-fig5}, an interferometric detector, such as {\sl LIGO},
has a wider sensitivity band, a few 100 Hz.

Wider bandwidth is obviously a significant improvement, as more information
from the source is made available with it. Earth based detectors do however
face an insurmountable sensitivity barrier: it is the so called
\emph{seismic wall} ---see again right panel in figure~\ref{lobo-fig5}---
caused by the vibrations of the ground, which cannot be sufficiently
damped at frequencies below some 10--50~Hz~\cite{lobo-vsusp}. In addition,
going to lower frequencies requires stretching the interferometer
armlength, eventually to prohibitive values.

Both requirements ---no seismic wall and long armlengths--- find a
natural solution in a space-borne GW antenna. This is the idea behind
the \lisa Project.

\section{\textsl{LISA}}
\label{lobo-sec3}

\lisa (Laser Interferometer Space Antenna) is a joint \esa--\nasa
Project. It was formally approved by \esa's Science Programme Committee
in November~2003 as a Fundamental Physics mission to fly in 2012. The
Project was also favourably informed by an independent Technology
Readiness and Implementation Plan assessment panel in April~2003 for
\nasa. Financial contribution to \lisa is currently envisaged as 50\,\%
from each side of the Atlantic.

The idea of placing a GW detector in space dates back to the mid 1980's,
and first designs were developed by various American scientists. None
of such proposals was however approved for actually flying in a \nasa
mission. In 1993 focus shifted to Europe, and after several studies
\lisa eventually became \emph{the} actual space-borne GW detector.
\nasa decided to join the Project in 1998~\cite{lobo-lfr}.

In the following sections we give a brief summary of the main design
and scientific characteristics of \lisa as they stand at the time of
writing. References will be provided for more detailed understanding
of the interested reader, and too technical material will be omitted.
In page \pageref{lobo-tab2} a short mission summary is also given.

\subsection{\textsl{LISA} concept}
\label{lobo-sec3.1}

\lisa is conceived as an interferometric detector to be sensitive to
GWs in the range of sub-milli-Hertz frequencies. As already discussed
in section~\ref{lobo-sec2.1}, this requires suitably long armlengths.
In the case of \lisa this will be five million kilometres, which sets
the resonance frequency in a few~mHz.

\begin{figure}[t!]
\centering
\includegraphics[width=9cm]{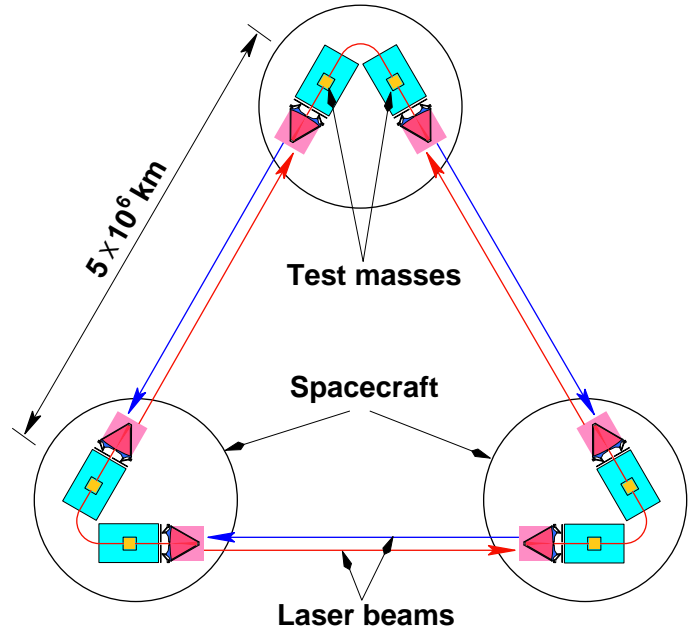}
\caption{Baseline design concept for \lisa.
\label{lobo-fig6}}
\end{figure}

The interferometer baseline configuration can be seen in figure~\ref{lobo-fig6}:
three spacecraft host two freely falling test masses each, and form an
equilateral triangle, 5\x\pten{6}~km to a side. Incoming GWs cause
distances between pairs of masses to change, and the changes are tracked
by interferometry between laser light beams travelling along the three
arms of the configuration. The interference generation scheme is however
not the usual one (e.g., Michelson) but rather an \emph{active} one,
based on \emph{transponders} instead of mirrors. Details willbe given
in section~\ref{lobo-sec3.4} below.

\begin{center}
\begin{table}[t!]
\caption{The assured galactic sources of \lisa.
\label{lobo-tab1}}
\medskip\medskip
\begin{tabular}{llccccc}
\hline\\[-1.2 em]
Class & Source & $d\/$/pc & $f$/mHz & $M_1/M_\odot$ & $M_2/M_\odot$ &
$h$/\pten{-22} \\
\hline\\[-1 em]
WD+WD  & WD\,0957-666     & 100  & 0.38 & 0.37 & 0.32    & 4 \\
       & WD\,1101+364     & 100  & 0.16 & 0.31 & 0.36    & 2 \\
       & WD\,1704+481     & 100  & 0.16 & 0.39 & 0.56    & 4 \\
       & WD\,2331+290     & 100  & 0.14 & 0.39 & $>$0.32 & $>$2 \\[1 em]
WD+sdB & KPD\,0422+4521   & 100  & 0.26 & 0.51 & 0.53    & 6 \\
       & KPD\,1930+2752   & 100  & 0.24 & 0.50 & 0.97    & 10 \\[1 em]
Am CVn & RXJ\,0806.3+1527 & 300  & 6.2  & 0.4  & 0.12    & 4 \\
       & RXJ\,1914+245    & 100  & 3.5  & 0.6  & 0.07    & 6 \\
       & KUV\,05184-0939  & 1000 & 3.2  & 0.7  & .092    & 0.9 \\
       & AM\,CVn          & 100  & 1.94 & 0.5  & .033    & 2 \\
       & HP\,Lib          & 100  & 1.79 & 0.6  & 0.03    & 2 \\
       & CR\,Boo          & 100  & 1.36 & 0.6  & 0.02    & 1 \\
       & V803\,Cen        & 100  & 1.24 & 0.6  & 0.02    & 1 \\
       & CP\,Eri          & 200  & 1.16 & 0.6  & 0.02    & 0.4 \\
       & GP\,Com          & 200  & 0.72 & 0.5  & 0.02    & 0.3 \\[1 em]
LMXB   & 4U\,1820-30      & 8100 & 3.0  & 1.4  & $<$0.1  & 0.2 \\
       & 4U\,1620-67      & 8000 & 0.79 & 1.4  & $<$0.03 & .06 \\[1 em]
W\,UMa & CC\,Com          & 90  & 0.105& 0.7  & 0.7      & 6 \\
\hline
\end{tabular}
\end{table}
\end{center}
\subsection{Scientific goals of \textsl{LISA}}
\label{lobo-sec3.2}

The sensitivity of \lisa is limited by various sources of noise which
happen in the frequency band around 1~mHz. This is represented in
figure~\ref{lobo-fig2}, left curve. In the lower part of the spectrum,
sensitivity is limited by signal loss because the resonant armlength
exceeds the interferometer's, and also because of internal sources
of noise associated with various measuring instruments, temperature
gradients, etc. In the resonance region, \emph{shot noise} is the
main source of disturbances: interferometric measurements are made
by means of photodiodes, and these are subject to \emph{quantum
fluctuations} which tend to mask low level signals. Finally, at
higher frequencies signals periodically cancel out due to nodes
in the interferometer response ---see equation~(\ref{lobo-eq2.6})---,
and shot noise increases, too.

If the forseen sensitivity is reached, \lisa will be able to see a number
of very interesting sources, such as mergers of large balck holes or even
GW backgrounds, see~\cite{lobo-bfs99} for a rather complete enumeration.
Unfortunately, galactic binaries are estimated to produce a background
GW luminosity in a part of \lisa's sensitivity band which will tend to
conceal individual events, unless they are very bright. This is
represented in figure~\ref{lobo-fig2} by the shaded area labeled
``unresolved galactic binaries'', and it can be compared to the sun's
day lihgt, which prevents the observation of stars and planets in the
optical frequency band. Even so, a number of galactic sources, whose
properties are known from electromagentic observation, are actually
guaranteed to be seen by \lisa if it works as planned. An updated list
of such sources is shown in table~\ref{lobo-tab1}. They fall in the area
labeled ``resolved galactic binaries'' in figure~\ref{lobo-fig2}.

\subsection{Orbits}
\label{lobo-sec3.3}

A crucial requirement for \lisa to work is of course that the test masses
be in \emph{free fall}. Within the solar system, this means that the
three spececraft should follow a geodesic motion determined by the
gravitational field of the Sun and the planets. At the same time,
such motion must be sufficiently stationary to ensure that relative
distances between spacecraft do not undergo variations which could
be wrongly taken as caused by real GWs, or which may conceal them.

\begin{figure}[t!]
\centering
\includegraphics[width=10.6 cm]{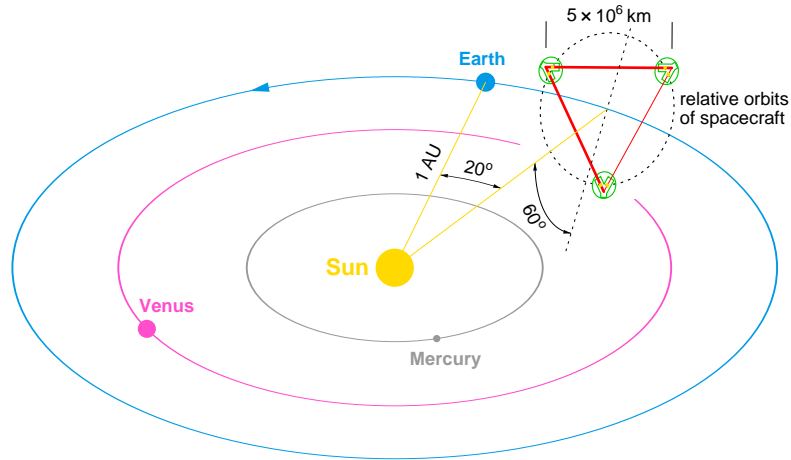}
\caption{Orbital placement of \lisa.
\label{lobo-fig7}}
\end{figure}

After some initial debate, it was agreed that \lisa should be placed
in a heliocentric orbit, rather than geocentric. It is however a
non-trivial problem to find a flight configuration which meets the
requirements of stationarity and suitability for \lisa. Analyses by
Bender, Folkner and others~\cite{lobo-billf,lobo-pete} resulted in
a series of remarkable conclusions which are schematically summarised
in figure~\ref{lobo-fig7}.

The baseline configuration is, as already stated, an equilateral triangle
5\x\pten{6}~km to a side. The barycentre of the triangle follows the
ecliptic, and the plane of the triangle is inclined 60$^\circ$ with
respect to the ecliptic plane at any given time; however its normal
does not point to a fixed position in the sky, but rather describes
a cone around the normal to the ecliptic, a complete tour per year.
Each of the three satellites revolves around the Sun in a nearly
circular orbit of eccentricity~0.01, inclined 1$^\circ$ relative
to the ecliptic, while the triangle itself rotates counterclockwise
about its barycentre, also with a period of one year. Finally, the
barycentre trails 20$^\circ$ behind the Earth, a figure which, unlike
the just quoted astrometric parameters, is not dynamically fixed, but
a compromise between telemetry constraints (satellites not too far)
and Earth-Moon perturbations (satellites not too near).

The reader may wonder at this stage about the \emph{stationarity} of
this configuration, the more so given that \lisa is supposed to perform
nano-meter laser interferometry: is it even conceivable that the
spacecraft keep their relative distances to within \pten{-9}~metres?

\begin{figure}[t!]
\centering
\includegraphics[width=12.5cm]{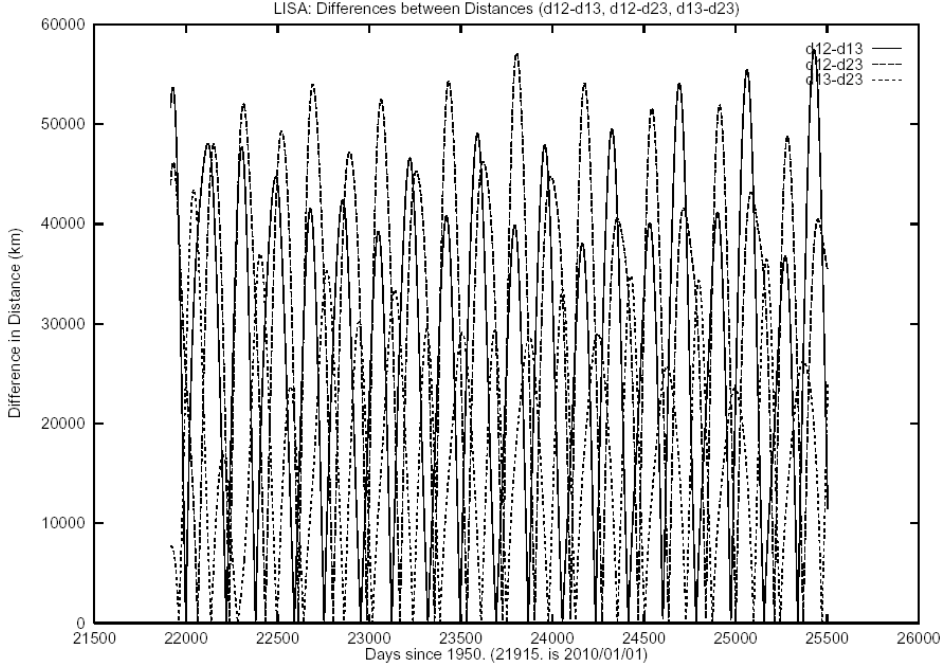}
\caption{Variations of the relative distances between the three pairs
of spacecraft of \lisa. Estimates correspond to a period of ten years
from January 1st 2010, around the time of flight of the mission. As
can be seen, maximum variations are about 50\,000~km, only a 10\,\%
of the nominal armlength of 5\x\pten{6}~km.
\label{lobo-fig8}}
\end{figure}

The answer to this question is obviously ``no'', but it must be admitted
that such question does not really address a relevant issue for \lisa\ldots\ 
Figure~\ref{lobo-fig8} shows how relative distances between the three
satellites vary over a period of 10 years, the (extended) scheduled
mission lifetime, from January 1st 2010 onwards. Dates of course affect
calculations, due to solar system ephemeris. The plot shows that distance
variations can be as large as 50\,000~km, 10\,\% of the baseline armlength
of 5\x\pten{6}~km. The relevant datum however is that variation periods
are in the order of a few months, well below the frequency sensitivity
range of \lisa, which corresponds to periods of only hours. This means
that \emph{fluctuations} of the latter period happen against a comparatively
very stable background baseline distance, which is well known at any given
time. In summary, if \lisa is quiet enough at \pten{-4}--\pten{-3}~Hz then
the possibility of sensing GWs in that frequency band may not be disturbed
by solar system induced armlength fluctuations.

Placement of the satellites in the inital coordinates and orientations
is aided by star trackers, working in conjunction with detailed star
databases. The laser itself is of course an essential pointing resource.
Micro thrusters take care of final adjustment.

\subsection{Interferometry}
\label{lobo-sec3.4}

The interferometry in \lisa has some differences with that in earth based
GW detectors, which are essentially Michelson, the most important being
related to the way light beams are made to interfere.

In \lisa, like in earth based detectors, the laser source is a Yag
Neodimium laser head, 1~watt effective power, which emits in a
stable infrared wavelength of 1.064~microns. The beam divergence is
small, 4\x\pten{-6} radians. Nevertheless in the case of \lisa the
light must travel 6 million kilometres before it reaches the other
end mass, which means the light power is spread over a circular spot
of about 20~km in diameter at that distance. Only a small fraction of
this power is collected by the receiver, actually \pten{-10}~watts.
If this were reflected back to the emitter's location then only some
200~photons per hour would be available for interference\ldots\ This
is by far much less than the photon counting noise in the sensitive
photodiodes, so the experiment would be impracticable.

The solution to this difficulty is the use of \emph{active} mirrors,
or \emph{transponders}: in the receiving satellite a new laser source
is installed which is \emph{phase locked} to the incoming light, and
shone back to the origin. On reception of two such signals from the
two remote satellites, a phase comparison (interference) is possible
which reveals any relative displacements between the test masses.

In practice, the laser light is modulated by means of acousto-optical
modulators (AOM). These are non-linear devices which alter the frequency
of the light in a controlled way, shifting it by some 100~MHz. The main
beam is split into two, and these are frequency shifted by two different
AOMs to frequencies which differ by about 1~kHz, the so called
\emph{heterodyne} frequency. In the end the signal will be sought
in the heterodyne modulated component, so that the valuable information
signal can be analysed at low frequencies ---see~\cite{lobo-gh} for a
thorough description.

\subsection{Inertial sensors and \textit{Drag free}}
\label{lobo-sec3.5}

A most delicate problem in \lisa is to make sure that the test masses do
actually follow geodesic motions to very high precision. Interplanetary
space is in fact a considerably hostile medium: solar radiation pressure,
ionising particle fluxes, and environmental magnetic fields are among the
agents which would perturb geodesic motion should the test masses be
floating unshielded in their orbits.

Apart from hosting the measuring and control instrumentation, the
spacecratf play a fundamental role in providing the test masses adequate
protection against external agents. Test masses are freely floating
\emph{inside} the spacecraft, and it is therefore the latter which
receive the impact of perturbations, eventually being driven away from
their geodesics. In order not to drag along the test masses with it, a
so called \emph{drag free} system is implemented in the satellites: this
consists in an \emph{inertial sensor} and an associated \emph{actuation
system}.

Each test masses is housed in a box whose walls are metallic plates which
form capacitors with the faces of the test mass itself. In equilibrium
conditions the mass is centred in the housing, and deviations thereof
result in capacity changes, which are detected by corresponding bias
voltage variations~\cite{lobo-billdf}. This is called \emph{inertial
sensor}~(IS). Its error signal is then used to send suitable ignition
commands to a set of {\sl FEEP} (Field Emission Electric Propulsion)
micro-thrusters which restore the centred position of the test masses
by acting on the spacecraft only. The {\sl FEEP\/}s produce very delicate
micro-newton forces by the ejection of ions, accelerated in a several
kilovolt electric field~\cite{lobo-feep}.

\subsection{\textsl{LISA} mission summary}
\label{lobo-sec3.6}

\begin{figure}[b!]
\centering
\includegraphics[width=6.2cm]{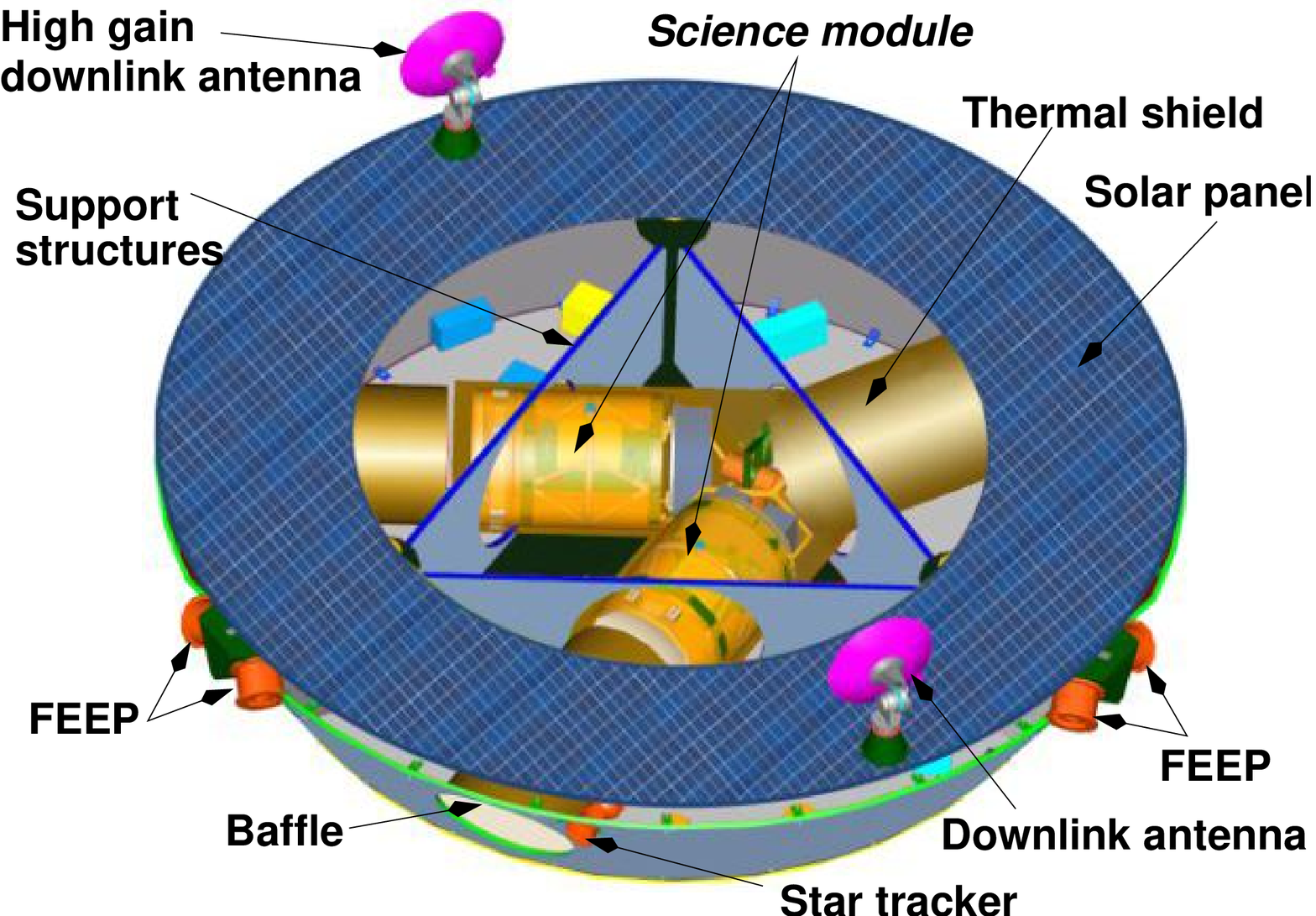}\hspace{1 em}
\includegraphics[width=6.2cm]{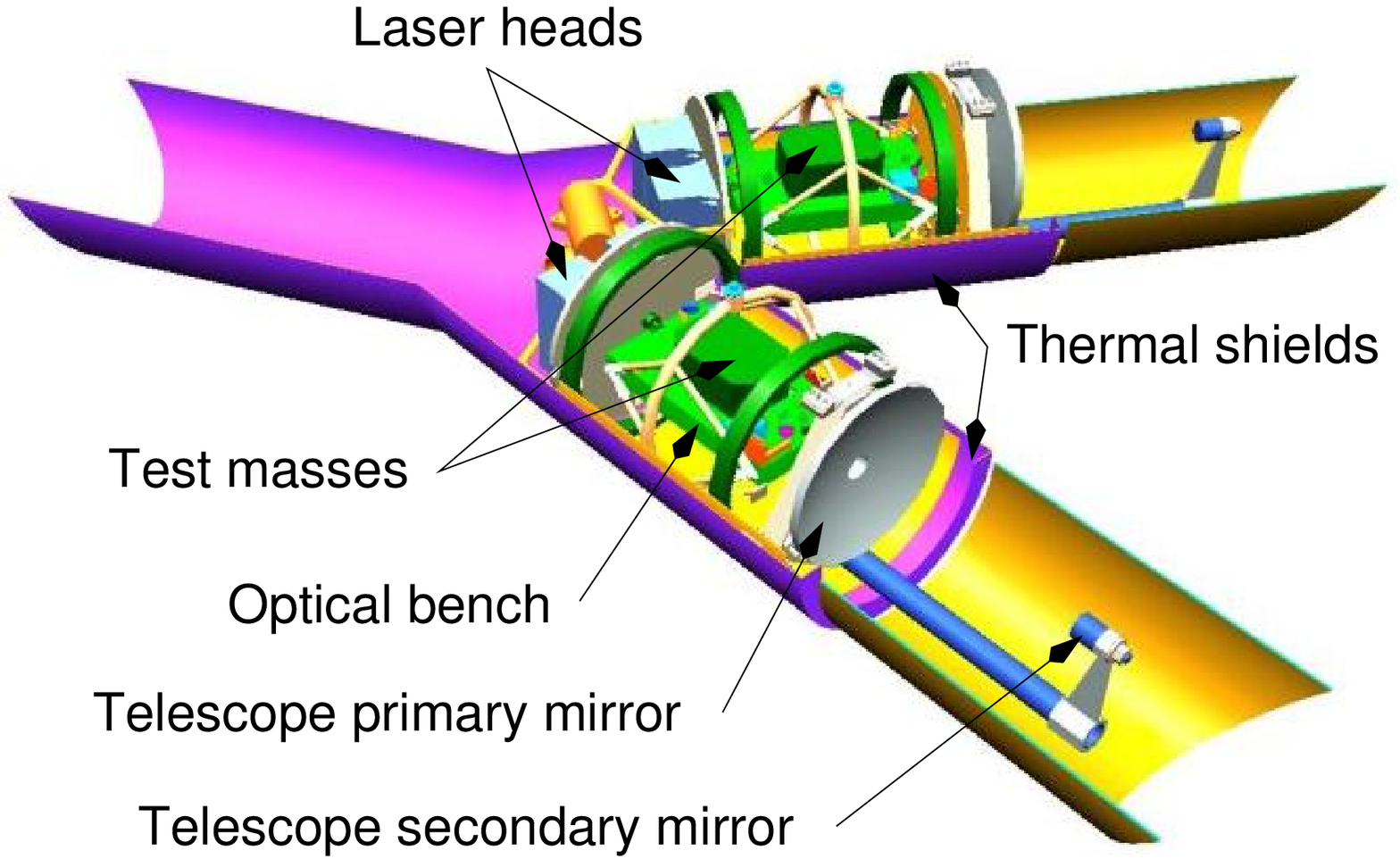}
\caption{Left: view of each of the three \lisa spacecraft. The top lid
(solar panel) is shown cut to see the interior. Right: detail of the
main parts of the science module.
\label{lobo-fig9}}
\end{figure}

Figure~\ref{lobo-fig9} shows a few details of the structure of the spacecraft
and the science module it hosts. Thermal stability is a major requirement,
as thermal gradients affect both the interferometry and the IS. The solar
panels partly stabilise the temperature, which should be about 300~K,
but additional thermal shields are necessary for fine tuning. Telescopes
are needed both for sending light to remote spacecraft, and to collect
incoming light from them; they are Cassegrain type, with a parabollic
primary mirror of 300~mm diameter and a secondary of 32~mm. 

The science module is a {\sf Y}-shaped tube with (obviously) a separation
of 60$^\circ$, and it includes all the elements described above. In order
to prevent sunlight from getting into the system, the ends of the {\sf Y}
carry 30$^\circ$ baffles. Aligned with them star trackers are attached
for coarse pointing operations. All together each spacecraft has a
diameter of 3~metres, and weighs 274~kg.

The table in page \pageref{lobo-tab2} presents a summary of various
parameters of the mission.

\begin{table}[p]
\label{lobo-tab2}
\small
\tabcolsep.45em
\vspace{-.5\baselineskip}
\begin{tabular}{|p{9em}|p{28.9em}|}\hline
  \multicolumn{2}{|c|}
  {\rule{0mm}{3.5ex}\large\bf \lisa Mission Summary} \\[.5ex] \hline\hline
Objectives: & Detection of low-frequency ($10^{-4}$ to $10^{-1}$~Hz)
              gravitational radiation with a strain sensitivity of
              $4\x 10^{-21}/\sqrt{\rm{Hz}}$ at 1\,mHz. \\
  & Galactic binaries (neutron stars, white dwarfs, etc.);
    extra-galactic targets are supermassive black hole binaries
    ({SMBH}-{SMBH} and {BH}-{SMBH}),
     {SMBH} formation, and   cosmic background GWs. \\ \hline
Payload:
  & Laser interferometry on six-degree-of-freedom capacitive sensing
    drag-free reference mirrors, housed in 3 spacecraft;
    arm lengths $5\x 10^6\,$km. \\
  & Each spacecraft has two lasers (plus two spares) which operate
    in a phase-locked transponder scheme. \\
  & Diode-pumped {Nd:YAG} lasers:~wavelength 1.064\,$\mu$m, output
    power 1\,W, Fabry-Perot cavity for frequency-stability
    of $\rm 30\,Hz/\sqrt{Hz}$. \\
  & Quadrant photodiode detectors with interferometer fringe resolution,
    corresponding to $4\x 10^{-5} \lambda/\sqrt{\rm Hz}$. \\
  & 30\,cm diameter f/1 Cassegrain telescope (transmit/receive),
    $\lambda/10$ outgoing wavefront quality. \\
  & Drag-free proof mass (mirror): 40 mm cube, Au-Pt alloy of low
    magnetic susceptibility ($<10^{-6}$);
    Titanium housing at vacuum $<10^{-6}$ Pa. \\ \hline
Orbit:
  & Each spacecraft orbits the Sun at 1 {AU}. \\
  & Spacecraft distributed at three vertices,
    defining an equilateral  triangle
    with a side length of $5\x 10^6$\,km (interferometer baseline). \\
  & Triangle rotates about its centre once a year. Centre follows
    ecliptic, $20^\circ$ behind the Earth. \\
Launcher:
  & Delta\,II\,7925\,H, 10 ft fairing, housing a stack of three
    \emph{composites} consisting of one science and one propulsion
    module each.\\
  & Each spacecraft has its own jettisonable propulsion
    module to provide a $\Delta V$ of 1300 m/s
    using solar-electric propulsion. \\ \hline
Spacecraft:                 & 3-axis stabilized drag-free spacecraft (three) \\
  \ *mass:              & 274 kg each spacecraft in orbit \\
  \ *propulsion module: & 142 kg one module per spacecraft \\
  \ *propellant:        & 22 kg for each propulsion module \\
  \ *total launch mass: & 1380 kg \\
  \ *power:             & 940 W each composite during cruise \\
  \ *power:             & 315 W each spacecraft in orbit \\
Drag-free:              & $3\x 10^{-15}\ \rm m/s^2$ (rms) in the band
                              \pten{-4} to $3\x10^{-3}$ Hz, achieved with
                              $6\x 4$ Cs or In {FEEP} thrusters \\
Pointing:               & few nano-rad/$\rm\sqrt{Hz}$ in the band
                                    \pten{-4} Hz to 1 Hz \\
Payload mass/power:     & 70 kg/72 W \emph{each} spacecraft \\
Science data rate:      & 672 bps \emph{all three} spacecraft \\
Telemetry:              & 7 kbps for about 9 hours inside two days \\ \hline
Mission Lifetime:       & 2 years (nominal), 10 years (extended)
  \\ \hline
\end{tabular}
\end{table}

\section{The \textsl{LISA Pathfinder} mission}

\lisa technological requirements are extremely demanding indeed. One
of the greatest difficulties to get it working is to ensure that the
\emph{drag free} subsystem works properly. Drag free systems have
been used in many missions since the 1960's, when they were first
introduced to compensate for height loss in low geocentric orbit
satellites, caused by rarified atmospheric gas friction ---hence
the name \emph{drag free}. Improvement has been enormous since,
but the best technology tested so far is still orders of magnitude
away from the needs of \lisa.

Because of the high risk of launching a very expensive mission whose
technology is not sufficiently verified, \esa decided that a previous,
smaller mission should be flown first to ensure proper technology readiness.
This mission is {\sl SMART-2\/} (Small Mission for Advanced Research
Technology number 2), now renamed as \lisa\ {\sl Pathfinder} (\lpf)\footnote{
Initially, {\sl SMART-2\/} was planned to also fly a technology test for
the mission {\sl DARWIN}, but this was dropped and {\sl SMART-2\/} is
currently a \lisa only mission.}. \lpf will carry on board the \ltp
(\lisa Test-flight Package), the payload which incorporates the
technology to be tested.

\lpf is scheduled for launch in late 2007, five years before \lisa
---provided of course the test is successful. This is a technological
mission, so it is not intended to measure any GWs. Its very high precision
distance measurements, however, have already prompted ideas to consider
potential applications of the analysis of its data stream~\cite{lobo-gazta}.
In the following sections we give a summary description of the main
elements and functions of the \ltp and the test.

\subsection{\textsl{LPF} mission concept}
\label{lobo-sec4.1}

One of the most demanding requirements of \lisa, as already stressed
in section~\ref{lobo-sec3.5}, is the \emph{drag free} subsystem. In
terms of spectral density of noise, the \lisa goal sensitivity is
\begin{equation}
 S^{1/2}_h(\omega)\leq 4\times\pten{-21}\,\left[
 1 + \fparen{f}{3\ {\rm mHz}}^{\!\!2}\right]\ {\rm Hz}^{-1/2}\ ,\quad
 0.1\ {\rm mHz}\leq\frac{\omega}{2\pi}\leq 0.1\ {\rm Hz}
  \label{lobo-eq4.1}
\end{equation}

This can be easily translated into relative acceleration noise by means
of the following considerations. Like in \eq{lobo-eq2.3}, the GW relative
acceleration between two test masses is given in terms of the amplitude
$h(t)$ by $\Delta a_{\rm GW}$\,=\,$L\,\ddot h/2$, where $L\/$ is the
rest distance between them. If non-GW forces, $\Delta F$, are also
present ---essentially non-wished, random perturbations--- then the
total acceleration is, clearly,
\begin{equation}
 \Delta a\equiv\frac{d^2\Delta x}{d t^2} = L\,\frac{d^2h}{dt^2}
 + \frac{\Delta F}{m}
 \label{lobo-eq4.2}
\end{equation}
where $m\/$ is the mass of one test mass. It is expedient to rewrite
this expression in frequency domain:
\begin{equation}
 \widetilde{\Delta a}(\omega) = -\omega^2\,\widetilde{\Delta x}(\omega) =
 -L\omega^2\,\widetilde{h}(\omega) + \frac{\widetilde{\Delta F}(\omega)}{m}
 \label{lobo-eq4.3}
\end{equation}
where a \emph{tilde} (\,$\tilde{}$\,) stands for Fourier transform.
If we now take spectral densities in the last equation we find,
assuming of course that the the true GW signal is deterministic,
that the spurious forces $\Delta F$ actually \emph{fake} a GW noise
with equivalent spectral density (rms)
\begin{equation}
 S_h^{1/2}(\omega) = \frac{S_{\Delta F}^{1/2}(\omega)}{mL\omega^2}\ ,
 \qquad {\rm equivalent\ signal}
 \label{lobo-eq4.4}
\end{equation}

In terms of acceleration noise of the test masses, the top level
requirement for \lisa, equation \eq{lobo-eq4.1}, can be rewritten as
\begin{equation}
 S^{1/2}_a(\omega)\leq 3\times\pten{-15}\,\left[
 1 + \fparen{f}{3\ {\rm mHz}}^{\!\!2}\right]\ 
 \frac{\rm m}{{\rm s}^2\,\sqrt{\rm Hz}}
 \ ,\quad 0.1\ {\rm mHz}\leq\frac{\omega}{2\pi}\leq
 0.1\ {\rm Hz}\,,\quad\lisa
 \label{lobo-eq4.5}
\end{equation}

\begin{figure}[b!]
\centering
\includegraphics[width=11cm]{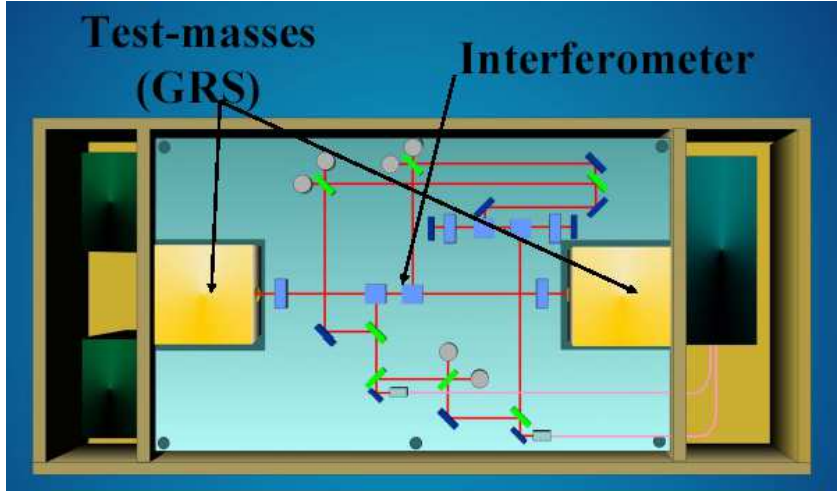}
\caption{Conceptual diagramme of \ltp. {\sl GRS} stands for
Gravitational Reference Sensor, or Inertial Sensor in the
jargon of this paper.}
\label{lobo-fig10}
\end{figure}

The \ltp concept is to squeeze two \lisa test masses into a small
satellite and check for drag free performance between these
---figure~\ref{lobo-fig10}: the 5\x\pten{6}~km of a \lisa arm is
thus compressed to a much shorter distance of some 30~centimetres.
It has been agreed that performance of \ltp can be considered
satisfactory if the requirement for \lisa is relaxed by one order
of magnitude, both in spectral density of noise and in frequency
band. More specifically~\cite{lobo-ltp-1.3},
\begin{equation}
 S^{1/2}_a(\omega)\leq 3\times\pten{-14}\,\left[
 1 + \fparen{f}{3\ {\rm mHz}}^{\!\!2}\right]\ 
 \frac{\rm m}{{\rm s}^2\,\sqrt{\rm Hz}}
 \ ,\quad 1\ {\rm mHz}\leq\frac{\omega}{2\pi}\leq
 30\ {\rm mHz}\,,\quad\ltp
 \label{lobo-eq4.6}
\end{equation}

\subsection{Philosophy of the test}
\label{lobo-sec4.2}

The instrumentation and design of the \ltp must ensure that any
\emph{residual accelerations}, i.e., those of unknown physical origin,
be below the requirement expressed by equation~\eq{lobo-eq4.6}. This
requires in turn a detailed \emph{apportioning} of different
contributions to the background noise.

An essential fact in this respect is the following: certain perturbing
agents \emph{couple} the spacecraft structure to the test masses, while
others are independent of them. A sort of master equation can thus be
set up:
\begin{equation}
 a_{\rm noise} = \frac{F_{\rm int}}{m} +
 \omega_{\rm p}^2\,\underbrace{\left(x_{\rm n} +
 \frac{F_{\rm S/C}}{M\omega_{\rm fb}^2}
 \right)}_{\mbox{\scriptsize\parbox[t]{22.5 ex}{
 $x_{\rm S/C}={\rm S/C}\rightleftharpoons{\rm TM}$
 {\rm relative\ distance}}}}
 \label{lobo-eq4.7}
\end{equation}
with the symbols meaning:

\begin{itemize}
 \setlength{\itemsep}{-0.4 ex}   
 \item $F_{\rm int}$: Total random force acting on a given test mass,
  independent of its position at any given time.
 \item $\omega_{\rm p}^2$: Elastic constant (or \emph{stiffness}) of
  coupling between the test mass and the spacecraft. This may be a
  negative number, if the equilibrium position is unstable relative
  to the considered deviation.
 \item $x_{\rm n}$: Random displacement fluctuations of the test mass.
 \item $F_{\rm S/C}$: Random force fluctuations acting on the spacecraft
  which back act on the test masses through the drag free actuation.
 \item $\omega_{\rm fb}^2$: Elastic constant of the above coupling.
  This is of course related to the response time of the actuation system.
 \item $M\/$ is the spacecraft mass, and $m\/$ is the test mass mass.
\end{itemize}

All the parameters in~\eq{lobo-eq4.7} must be evluated on the basis of
experimental measurement, and to this end several measuring runs and
operation modes have been designed~\cite{lobo-tlsr}. In each case,
the \ltp interferometer will be used as a \emph{diagnostic} instrument,
as its forseen sensitivity is largely sufficient to meet the measurement
demands necessary to establish the validity of the limit set by
equation~\eq{lobo-eq4.6}. The description of further technical details
is beyond the scope of this article, but information can be found in the
References section below.

\subsection{\textsl{LPF} mission summary}
\label{lobo-sec4.3}

\begin{figure}[t!]
\centering
\includegraphics[width=6.2cm]{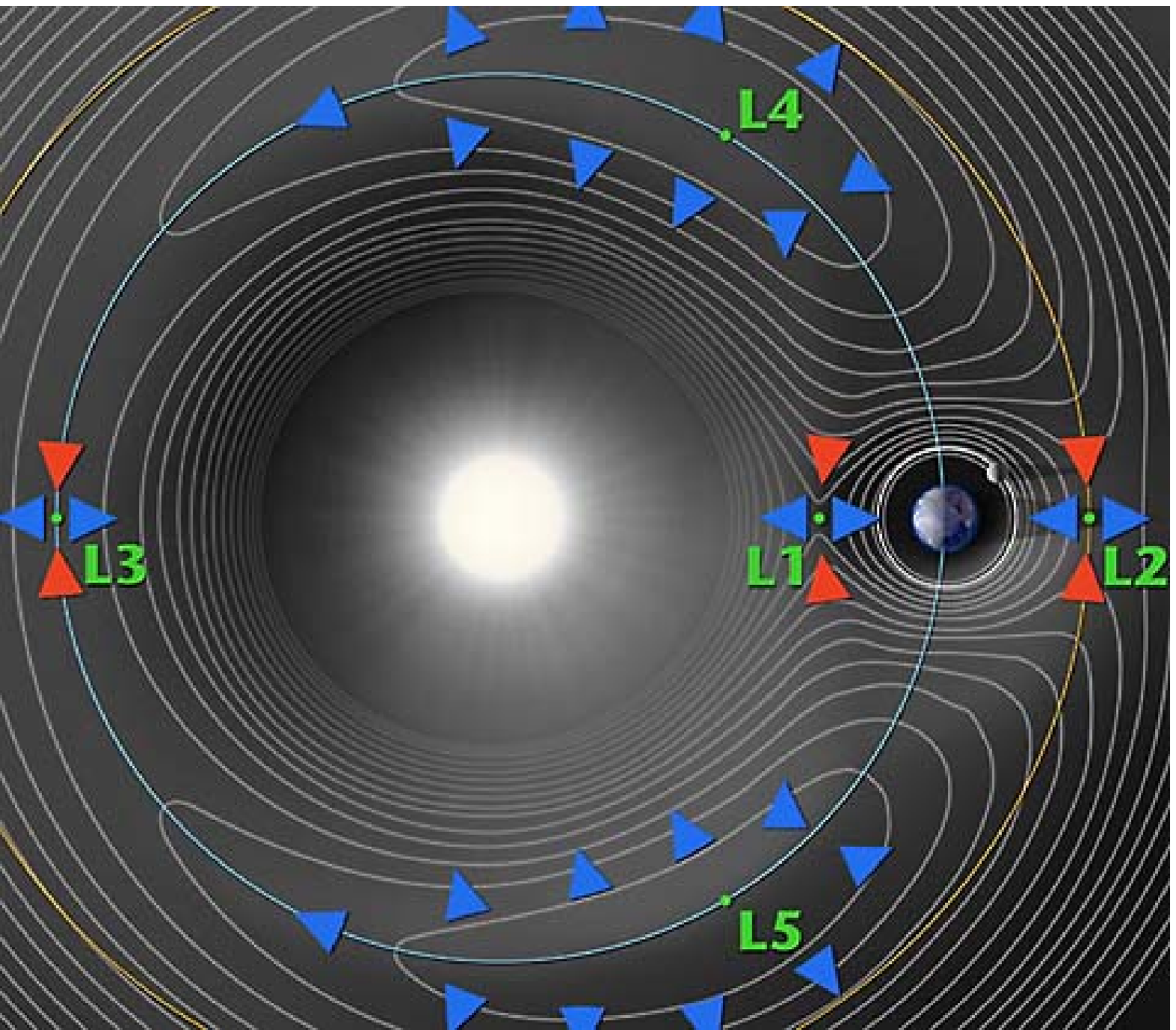}\hspace{1 em}
\includegraphics[width=6.2cm]{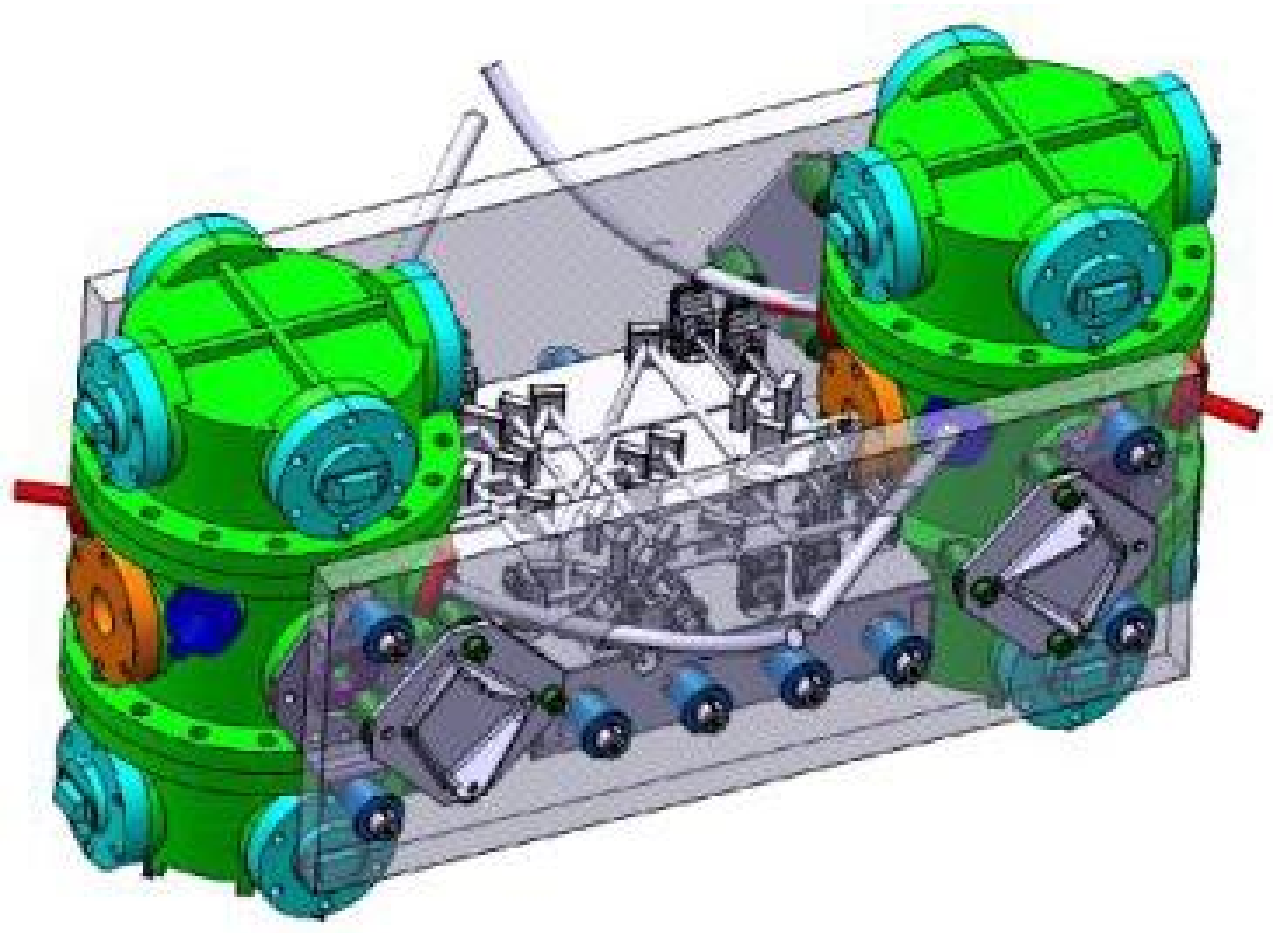}
\caption{Left: the Lagrange points of the Earth-Sun system; \lpf will
go to \emph{L1}. Right: engineering schematics of the \ltp; the
cylindrical towers on the ends host the inertial sensors, and the
platform in between is the optical bench.
\label{lobo-fig11}}
\end{figure}

The \lpf mission will be placed in a Lissajous orbit around the Lagrange
point \emph{L1} of the Earth-Sun system, some 1.5~million kilometres from
here ---figure~\ref{lobo-fig11}, left. Launch date is forseen for late 2007,
and it will take the satellite some three months to reach the operations
destination. The various \ltp tests are scheduled to last 100 days, 23~of
which will be in joint operation with {\sl DRS\/} (Disturbance Reduction
System)\footnote{
{\sl DRS\/} is a second technological payload, indepenently developed
for \lisa by \nasa scientists and engineers. It will also be flown on
board \lpf, an \esa alone mission.}.
The {\sl DRS\/} will require another 100 day period of tests, so the
whole mission lifespan is about a half year.

Some of the internals of the \ltp core are shown in figure~\ref{lobo-fig11},
right. The green towers on either side host the inertial sensors and
the \emph{caging mechanism}, a mechanical system which holds the test
masses fixed during launch and then releases them for operation. Optical
fibres (in red) are connected to the towers to let ultraviolet light get
into the IS for eventual electrical discharge by photoelectric effect:
this is necessary as cosmic rays and solar flares diposit charge in
the IS and degrade its performance. Between the towers lies the
optical bench with all its mirrors, beam-spliters and photodiodes,
necessary for the various interferometric measurements. The optical
bench is manufactured with a low thermal coefficient of dilatation
to ensure stability, and the optical elements are \emph{bonded} to
it rather than glued, for the same reason. The optical bench is tied
to the ISs by means of vertical glass panels, as shown.

The entire apparatus is placed inside a double layer thermal shield
---not shown in the figure--- which affords thermal insulation and
a mechanical interface with the spcecraft. The laser head is placed
outside the latter, but inside the spacecraft, and the light is fed
in by menas of optical fibres, too. The electronics boxes are placed
outside as well to reduce heating and interference effects.

\subsection{The Spanish role in \textsl{LPF}}
\label{lobo-sec4.4}

An important part of the \ltp is the so called \emph{diagnostics
subsystem}. This consists in a number of sensors which are meant to
monitor various perturbations affecting the scientifically relevant
measurements. More specifically, the diagnostics subsystem must
monitor

\begin{itemize}
\setlength{\itemsep}{-0.4 ex}   
 \item Thermal gradients across the IS
 \item Electric charge accumulation in the tsts masses due to the
	impact of ionising particles in cosmic rays and solar flares
 \item Magnetic field gradients across the IS
 \item Solar radiation pressure fluctuations
\end{itemize}

The appropriate sensors need to be programmed and controlled, and
this is done by means of the Data Management Unit ({\sl DMU\/}),
which also hosts various service electronics, and post-processes
the main science data stream coming from the interferometer.

The international consortium of the \ltp offered the responsibility
of the design, manufacture and integration of both the diagnostics
subsystem and the {\sl DMU\/} to a Spanish group, which is in the
Institut d'Estudis Espacials de Catalunya ({\sl IEEC\/}), in
Barcelona. This happened in early 2003, after a larger collaboration
had been set up a year earlier to get into the scientific teams of
\lisa. That group includes {\sl IEEC\/}, two Universities from
Barcelona ({\sl UB\/} and {\sl UPC\/}), the Consejo Superior de
Investigaciones Cient\'\i ficas, and the Universities of Valencia
and Alicante. The collaboration is currently active and looking into
the longer term future of the GW detection Project. During 2003,
the Barcelona part of the team was consolidated as a technological
partner in \lpf, and has made industrial contacts to develop the
engineering parts of its assigned tasks. A significant project
proposal was submitted to the Spanish National Space Programme
({\sl PNE\/}) in December 2003, and at the time of writing the
report of the review process is underway. The contribution by
Miquel Nofrarias in this Proceedings book reports on progress
being made on the issue of thermal diagnostics.

\section{Conclusions}

The 21st century is witnessing an unprecedented scientific effort
aiming at the explicit detection of GWs. The first endeavours in
this direction date back to the 1960's, after pioneering work by
Joseph Weber. Progress has been enormous since, thanks to the
dedication of two generations of scientists. The new large
interferometric detectors ({\sl VIRGO}, {\sl LIGO}, {\sl TAMA},
{\sl GEO-600}) are close to reaching full performance, and they
could well see GWs for the first time within the next few years.
\lisa is scheduled to fly in 2012, so it fully joins the course
of events in this world-wide scientific programme.

There are however two very specific characterisitics of \lisa which
make a qualitative difference with earth based GW antennas: one is
its \emph{frequency range}, around one milli-Hertz, and the other
is that \lisa is \emph{guaranteed} the sighting of GW signals from
a score of galactic sources.

\lisa is an expensive and difficult Project, as shown in this article.
In particular, technology readiness is a real issue, and this is why
\lpf is programmed to fly first. Prospects however look good, as hard
work by many scientists, engineers and institutions continues to be done
with considerable devotion. This has resulted in unanimous approval of the
mission by \esa's Space Programme Committee in November 2003. In summary,
\lisa is a major challenge, but truly valuable science outcome from it
appears to be a sound prospect.

\end{document}